\documentclass[prd,aps,showpacs,nofootinbib,
eqsecnum]{revtex4}
\usepackage{amsmath}
\usepackage{amssymb}
\usepackage{amsfonts}
\usepackage{graphicx,bm}
\usepackage{dcolumn}
\usepackage{color,amsxtra}
\usepackage{epsf}
\usepackage{enumerate}
\usepackage{hhline}
\usepackage{array}
\usepackage{tabularx}
\newcommand{\be}{\begin{equation}}
\newcommand{\ee}{\end{equation}}
\newcommand{\bea}{\begin{eqnarray}}
\newcommand{\eea}{\end{eqnarray}}
\newcommand{\beaa}{\begin{eqnarray*}}
\newcommand{\eeaa}{\end{eqnarray*}}

\newcommand{\nn}{\nonumber \\}
\newcommand{\e}{\mathrm{e}}

\newcommand{\Eqn}[1]{&\hspace{-0.2em}#1\hspace{-0.2em}&}

\def\be{\begin{equation}}
\def\ee{\end{equation}}
\def\bea{\begin{eqnarray}}
\def\eea{\end{eqnarray}}

\def\nn{\nonumber \\}
\def\e{\mathrm{e}}

\begin{document}

\tolerance=5000

\title{Scalar Domain Wall as the Universe}

\author{Yuta Toyozato$^{1, }$\footnote{
E-mail address: toyozato@th.phys.nagoya-u.ac.jp}, 
Kazuharu Bamba$^{2, }$\footnote{
E-mail address: bamba@kmi.nagoya-u.ac.jp}, 
Shin'ichi Nojiri$^{1, 2,}$\footnote{E-mail address:
nojiri@phys.nagoya-u.ac.jp}
}

\affiliation{
$^1$ Department of Physics, Nagoya University, Nagoya
464-8602, Japan \\
$^2$ Kobayashi-Maskawa Institute for the Origin of Particles and
the Universe, Nagoya University, Nagoya 464-8602, Japan}

\begin{abstract}

We develop a formulation where for an arbitrarily given warp factor, we construct 
a scalar field action which reproduces the warp factor as a solution of the Einstein equation 
and field equation corresponding to the action. 
This formulation could be called as the reconstruction.    
By using the formulation of the reconstruction, we construct models which have an exact 
solution describing the domain wall. The shape of the domain wall can be flat, 
de Sitter space-time, or anti-de Sitter space-time.  
In the constructed domain wall solutions, there often appears ghost with negative kinetic energy. 
We give, however, an example of the de Sitter domain wall solution without ghost, 
which could be a toy model of inflation. We also investigate the localization of gravity as in the Randall-Sundrum model. 
It is demonstrated that the four dimensional 
Newton law could be reproduced even in the de Sitter space-time domain wall solution. 
We show that we can construct a space-time, where the domain wall is the general Friedmann-Robertson-Walker (FRW) 
universe and the warp factor can be arbitrary. 
For such a construction, we use two scalar fields. 
It is also illustrated that the scalar field equations are equivalent to the Bianchi identities: 
$\nabla^\mu \left( R_{\mu\nu} - \frac{1}{2}R g_{\mu\nu} \right)=0$.  

\end{abstract}

\pacs{95.36.+x, 98.80.Cq}

\maketitle

\section{Introduction \label{I}}

There are many scenarios that our universe could be a brane in the higher dimensional 
space-time \cite{Randall:1999ee,Randall:1999vf,Dvali:2000hr,Deffayet:2000uy,Deffayet:2001pu}. 
The inflationary brane world models were also considered by using the trace anomaly 
in \cite{Nojiri:2000eb,Hawking:2000bb,Nojiri:2000gb}. 
Before the brane world scenario, there was a scenario that we live 
on the domain wall \cite{Lukas:1998yy}, and 
bent domain wall \cite{Kaloper:1999sm} as well as 
dynamical domain wall \cite{Chamblin:1999ya} were also investigated. 
After that there were many activities in the domain wall or thick brane universe 
scenario \cite{DeWolfe:1999cp,Gremm:1999pj,Csaki:2000fc,Gremm:2000dj,Kobayashi:2001jd,Slatyer:2006un}. 
The domain wall has a thickness and the brane could be regarded as a limit 
that the thickness of the domain wall vanishes. 

In this paper, we consider the domain wall by using the scalar field. 
We construct models where the shape of the domain wall is flat, de Sitter space-time, 
or anti-de Sitter space-time. 
In case that the shape of the domain wall is de Sitter space-time, 
if we regard the domain wall as our universe, the de Sitter space-time 
may express inflation and/or the accelerating expansion of the current universe. 
It is developed a formulation to construct an action which reproduces 
an arbitrarily given warp factor as a solution of the Einstein equation 
and field equation corresponding to the action. 
This formulation could be called as the reconstruction.    
By using this formulation\footnote{
About the reconstruction for cosmology, see \cite{Nojiri:2010wj}.}
we give models which have an exact solution 
describing the domain wall. 
In the seminal paper \cite{DeWolfe:1999cp}, a formulation of the reconstruction has 
been proposed. In case of the brane with vanishing thickness, the potential which realizes 
the arbitrary warp factor has been analytically found. In case of the domain wall or thick 
brane, we need to solve first order differential equations to find the potential. 
In the formulation which we propose, even for the domain wall, we only need algebraic calculations 
and the action of the model can be 
obtained straightforwardly.  
We also show that there appears massless graviton propagating 
on the four dimensional domain wall. The four dimensional massless graviton is the zero 
mode of the five dimensional graviton and the existence of the four dimensional massless 
graviton may tell that the four dimensional gravity, where the Newton potential 
behaves as $1/r$ for the distance $r$, could be reproduced on the domain wall. 
Furthermore, we explore the models including two scalar fields and show that 
we can construct a space-time, where the domain wall is the general FRW 
universe and the warp factor can be arbitrary. 
The scalar field equations can be identified with the Bianchi identities: 
$\nabla^\mu \left( R_{\mu\nu} - \frac{1}{2}R g_{\mu\nu} \right)=0$ and therefore 
the equations can be satisfied automatically. 

In the next section, we explain the formulation to obtain models which admit 
exact solutions describing the domain wall and we also give some examples. 
In the domain wall solutions, there appears ghost in general. 
The ghost has negative kinetic energy. We give, however, 
an example of de Sitter domain wall without ghost, which can be a toy model of 
inflation. 
In Section \ref{III}, we investigate the localization of gravity and show the 
four dimensional Newton law could be reproduced. 
In Section \ref{IIIb}, we propose a two scalar model. 
By using the model, we show that we can construct a space-time, 
where the domain wall is the general FRW 
universe and the warp factor can be arbitrary. 
In Section V, we present examples of reconstructed model.
The last section is devoted to the summary and discussion. 

We use units of $k_\mathrm{B} = c = \hbar = 1$ and denote the
gravitational constant $8 \pi G$ by
${\kappa}^2 \equiv 8\pi/{M_{\mathrm{Pl}}}^2$
with the Planck mass of $M_{\mathrm{Pl}} = G^{-1/2} = 1.2 \times 10^{19}$GeV. 

\section{Constructing the action generating an exact domain wall solution \label{II}}

In this section, based on \cite{Bamba:2011nm}, we show how we can construct models 
which admit the exact solutions describing the domain wall. 
We use a procedure proposed in Ref.~\cite{Capozziello:2005tf}. 
This formulation is a kind 
of reconstruction, that is, for an arbitrary warp factor, we specify the action of 
the Einstein gravity coupled with a scalar field which has the solution corresponding to the scale factor. 

Our model action is as follows:
\be
\label{I1}
S = 
\int d^D x \sqrt{-g} \left( \frac{R}{2\kappa^2} - \frac{\omega(\varphi)}{2}
\partial_\mu \varphi \partial^\mu \varphi 
 - \mathcal{V} (\phi) \right)\, .
\ee
We 
assume the 
$D=d+1$ dimensional warped metric
\be
\label{I2}
ds^2 = dy^2 + L^2 \e^{u(y)} \sum_{\mu,\nu=0}^{d-1} {\hat g}_{\mu\nu} dx^\mu dx^\nu\, ,
\ee
with $L$ being a dimensionless constant. 
We 
suppose that the scalar field only depends on $y$. 
In the metric (\ref{I2}), 
${\hat g}_{\mu\nu}$ is the metric of the $d$-dimensional Einstein manifold 
defined by ${\hat R}_{\mu\nu} = \left[ \left (d-1\right) / l^2 \right] 
{\hat g}_{\mu\nu}$. The de Sitter space (the anti-de Sitter space) 
corresponds to $1/l^2>0$ ($1/l^2<0$), whereas 
the flat space to $1/l^2 = 0$. 
The $(y,y)$ and $(\mu,\nu)$ components of the Einstein equation are given by
\bea
\label{I3}
&&
 - \frac{d(d-1)}{2l^2} \e^{-u} + \frac{d(d-1)}{8} \left( u'\right)^2 
 = \frac{1}{2}\omega(\varphi) \left(\varphi'\right)^2 - \mathcal{V}(\varphi) \, , \\
\label{I4}
&&
 - \frac{(d-1)(d-2)}{2l^2} \e^{-u} + \frac{d-1}{2} u'' + \frac{d(d-1)}{8} \left( u'\right)^2 
 = - \frac{1}{2}\omega(\varphi) \left(\varphi'\right)^2 - \mathcal{V}(\varphi) \, .
\eea
Here the prime denotes the derivative with respect to $y$. 
Now $\varphi$ may be chosen as 
$\varphi=y$. 
Moreover, 
we 
take $\kappa^2 = 1$. 
Then Eqs.~(\ref{I3}) and (\ref{I4}) lead to 
\bea
\label{I5}
\omega(\varphi) \Eqn{=} - \frac{d-1}{2}u'' - \frac{d-1}{l^2} \e^{-u}\, , \\
\label{I6}
\mathcal{V}(\varphi) \Eqn{=} - \frac{d-1}{4} u'' - \frac{d(d-1)}{8} \left( u' \right)^2 
+ \frac{(d-1)^2}{2l^2}\e^{-u}\, .
\eea
Hence, the energy density $\rho$ is now described as \cite{Bamba:2011nm}
\be
\label{I7}
\rho = \frac{\omega(\varphi)}{2}\left( \varphi' \right)^2 + \mathcal{V}(\varphi)
= - \frac{d-1}{2} u'' - \frac{d(d-1)}{8} \left( u' \right)^2 + \frac{(d-1)(d-2)}{2l^2}\e^{-u}\, .
\ee
The pressure $p_y$ for the $y$ direction and the pressures $p$ for other spatial directions are different 
with each other and given by
\bea
\label{I7B}
&& p_y =  \frac{\omega(\varphi)}{2}\left( \varphi' \right)^2 - \mathcal{V}(\varphi)
= \frac{d(d-1)}{8} \left( u' \right)^2 - \frac{(d-1)(d-2)}{2l^2}\e^{-u}\, , \nn
&& p = \rho 
= - \frac{d-1}{2} u'' - \frac{d(d-1)}{8} \left( u' \right)^2 + \frac{(d-1)(d-2)}{2l^2}\e^{-u}\, .
\eea
Provided that 
the $D$ dimensional space is asymptotically flat, we see 
$u\to 0$ 
in the limit of $\left| y \right| \to \infty$, 
the second term becomes dominant 
in (\ref{I5}), 
$\omega(\varphi) \sim - (d-1) /l^2$ if $1/l^2\neq 0$. 
If $\omega(\varphi)$ is negative, 
which corresponds to the de Sitter space with $1/l^2>0$, the scalar field 
$\varphi$ is 
ghost. 
For $1/l^2=0$, we obtain 
$\omega(\varphi) = - (d-1)u''/2$. 
Thus we impose the 
$Z_2$ symmetry of the metric, which is the invariance under the transformation 
$y\to -y$. 
In addition, 
we 
suppose 
the $D$ dimensional space is asymptotically flat. 
In such a case, 
there must exist a region in which 
$\omega(\varphi)$ is negative and hence $\varphi$ is ghost. 
We should also remark that the energy density often becomes negative. 
In any case, 
if we allow 
the ghost and negative energy density, 
for an arbitrary $u$, 
we find a model which permits 
$u$ 
to be a solution of the Einstein equation. 
Furthermore,  
the problem of ghost can be avoided in case that the extra dimensions are compact. 

As an example, 
we may examine 
\be
\label{I8}
u = u_0 \e^{-y^2/y_0^2}\, ,
\ee
where $u_0$ and $y_0$ are constants. 
We explore  
the model
\bea
\label{I9}
\omega(\varphi) \Eqn{=} - (d-1)\left( \frac{2\varphi^2}{y_0^4} 
 - \frac{1}{y_0^2} \right)\e^{- \varphi^2/y_0^2} 
 - \frac{(d-1)}{l^2}\e^{-u_0 \e^{-\varphi^2/y_0^2}}\, ,\nn
\mathcal{V}(\varphi) \Eqn{=} - \frac{d-1}{2}\left( \frac{2\varphi^2}{y_0^4} 
 - \frac{1}{y_0^2} \right)\e^{- \varphi^2/y_0^2} 
+ \frac{(d-1)^2}{l^2}\e^{-u_0 \e^{-\varphi^2/y_0^2}}\, .
\eea
In this model, as a solution of the Einstein equation 
we acquire 
$u$ in (\ref{I8}). 
Furthermore, 
we find the following distribution of the 
energy density \cite{Bamba:2011nm}
\be
\label{I10}
\rho(y) = - \frac{d-1}{2}\left( \frac{2y^2}{y_0^4} - \frac{1}{y_0^2} \right)\e^{- y^2/y_0^2} 
+ \frac{(d-1)^2}{l^2}\e^{-u_0 \e^{-y^2/y_0^2}}\, . 
\ee
As a result, the energy density $\rho(y)$ is localized at $y\sim 0$ and 
therefore 
a domain wall is made. 
We also mention that for 
$1/l^2>0$, the shape of the domain wall is 
a de Sitter space, 
and hence it could 
represent the accelerating universe. 

\subsection{(Anti-)de Sitter space-time  \label{IIa}}

As a preparation to deal with the domain wall, here we consider the 
(anti-)de Sitter space-time 
and the flat space-time. These space-times do not 
correspond to brane or domain wall but we give explicit formula for later use.  

We now 
study the de Sitter space-time solution. 
When $L^2=l^2>0$ ($L$ appears in (\ref{I2})), if $u(y)$ is given by
\be
\label{dS1}
u = 2 \ln \cosh \frac{y}{l}\, ,
\ee
we find 
\be
\label{dS2}
\omega(\varphi) = 0 \, ,\quad 
\mathcal{V}(\varphi) = \frac{d(d-1)}{2l^2}\, .
\ee

We may also investigate 
the anti-de Sitter solution. 
When $L^2=l^2>0$, if $u(y)$ is given by 
\be
\label{AdS1}
u = 2\ln \sinh \frac{y}{l}\, ,
\ee
we acquire
\be
\label{AdS2}
\omega(\varphi) = 0 \, ,\quad 
\mathcal{V}(\varphi) = - \frac{d(d-1)}{2l^2}\, .
\ee
Here $y$ is restricted to be $y\geq 0$. 
On the other hand, when $l^2 = - {\tilde l}^2 = - L^2 <0$, 
if $u(y)$ is given by 
\be
\label{AdS3}
u = 2\ln \cosh \frac{y}{\tilde l}\, ,
\ee
we have 
\be
\label{AdS4}
\omega(\varphi) = 0 \, ,\quad 
\mathcal{V}(\varphi) = - \frac{d(d-1)}{2{\tilde l}^2}\, .
\ee
When $1/l^2=0$, 
if $u(y)$ is given by 
\be
\label{AdS5}
u = \frac{2y}{L}\, ,
\ee
where $L$ is a constant, we obtain 
\be
\label{AdS6}
\omega(\varphi) = 0 \, ,\quad 
\mathcal{V}(\varphi) = - \frac{d(d-1)}{2{L^2}}\, .
\ee

We now examine 
the flat space-time. 
When $l^2>0$, we find 
\be
\label{AdS7}
u= 2\ln \frac{y}{l}\, , \quad \omega(\varphi) = \mathcal{V}(\varphi) = 0\, .
\ee
Here $y\geq 0$. 
When $1/l^2=0$, we, of course, acquire
\be
\label{AdS8}
u = \omega(\varphi) = \mathcal{V}(\varphi) = 0\, .
\ee

\subsection{Randall-Sundrum like model  \label{IIb}}

In case of the second Randall-Sundrum model \cite{Randall:1999vf}, we have
\be
\label{RS1}
1/l^2 = 0\, ,\quad u(y) = - \frac{2|y|}{L}\, .
\ee
This model shows the localization of the gravity 
on the four dimensional brane.  
Motivated by 
the model (\ref{RS1}), we analyze 
the following model,
\be
\label{RS2}
1/l^2 = 0\, ,\quad u(y) = - \frac{2\sqrt{y^2 + y_0^2}}{L}\, .
\ee
Here $y_0$ is a constant and the Randall-Sundrum model (\ref{RS1}) corresponds 
to this model in the limit of $y_0\to 0$.
Then we obtain
\be
\label{RS3}
\omega(\varphi) = \frac{(d-1) y_0^2}{L \left(\varphi^2 + y_0^2\right)^{\frac{3}{2}}}\, ,\quad 
\mathcal{V}(\varphi) = \frac{(d-1) y_0^2}{2L \left(\varphi^2 + y_0^2\right)^{\frac{3}{2}}} 
 - \frac{d(d-1)\varphi^2}{2L^2 \left(\varphi^2 + y_0^2\right)}\, .
\ee
Then the energy density is given by
\be
\label{RS4}
\rho(y) = \frac{(d-1) y_0^2}{L \left(y^2 + y_0^2\right)^{\frac{3}{2}}} 
 - \frac{d(d-1)y^2}{2L^2 \left(y^2 + y_0^2\right)}\, .
\ee
In the limit of $y_0\to 0$, the second term in (\ref{RS4}) gives the 
negative cosmological constant corresponding to the anti-de Sitter space-time 
and the first term gives a $\delta$-function: 
\be
\label{RS5}
\frac{(d-1) y_0^2}{L \left(y^2 + y_0^2\right)^{\frac{3}{2}}} \to \frac{2(d-1)}{L}\delta (y)\, ,\quad 
 - \frac{d(d-1)y^2}{2L^2 \left(y^2 + y_0^2\right)} \to - \Lambda \equiv - \frac{d(d-1)}{2L^2} \, .
\ee
On the other hand, $p_y$ and $p$ in (\ref{I7B}) are given by
\be
\label{RS6}
p_y = \frac{d(d-1)y^2}{2L^2 \left(y^2 + y_0^2\right)} \to \Lambda = \frac{d(d-1)}{2L^2}\, ,\quad 
p = \rho \to \frac{2(d-1)}{L}\delta (y) - \frac{d(d-1)}{2L^2} \, .
\ee

\subsection{de Sitter domain wall and brane  \label{IIc}}

We now explore 
the de Sitter domain wall and brane with $l^2>0$, 
which may correspond to the expanding universe. 
The four dimensional de Sitter brane solution in five dimensional anti-de Sitter 
space-time using the trace anomaly is given by
\cite{Nojiri:2000eb,Hawking:2000bb,Nojiri:2000gb}. 

Motivated by 
(\ref{AdS1}), we may study 
\be
\label{dSb1}
u = 2\sinh \left( \frac{\sqrt{y_0^2 + z_0^2} - \sqrt{y_0^2 + z^2}}{l'} \right)\, , \quad 
l' \equiv \frac{l z_0}{\sqrt{y_0^2 + z_0^2}}\, .
\ee
Here $y_0$ and $z_0$ are positive constants and we assume $-z_0\leq z \leq z_0$. 
In the limit of $y_0\to 0$, we find 
\be
\label{dSb2}
\sqrt{y_0^2 + z_0^2} - \sqrt{y_0^2 + z^2} \to z_0 - |z|\, ,
\ee
which implies 
that the space-time given by (\ref{dSb1}) in this limit can be obtained as follows: 
First cut two anti-de Sitter spaces given by (\ref{AdS1}) at $y=z_0$. Second glue two space-times 
with the region $0 \leq y \leq z_0$ at $y=z_0$. Then in the limit (\ref{dSb2}), $z=0$ corresponds to $y=z_0$ 
and $z=\pm z_0$ to $y=0$. 
Then we find
\bea
\label{dSb3}
\omega(\varphi) &=& (d-1) \left( \frac{\frac{\varphi^2}{l^2 \left( \varphi^2 + y_0^2 \right)}}{
\sinh^2 \left( \frac{\sqrt{y_0^2 + z_0^2} - \sqrt{y_0^2 + \varphi^2}}{l'} \right)} 
 - \frac{y_0^2}{l' \left( \varphi^2 + y_0^2 \right)^{\frac{3}{2}}}
\frac{ \cosh \left( \frac{\sqrt{y_0^2 + z_0^2} - \sqrt{y_0^2 + \varphi^2}}{l'} \right)}{
\sinh \left( \frac{\sqrt{y_0^2 + z_0^2} - \sqrt{y_0^2 + \varphi^2}}{l'} \right)} \right) \nn
&& - \frac{d-1}{l^2 \sinh^2 \left( \frac{\sqrt{y_0^2 + z_0^2} - \sqrt{y_0^2 + \varphi^2}}{l'} \right)} 
\, , \nn
\mathcal{V}(\varphi) &=& \frac{(d-1)}{2}  \left( \frac{\frac{\varphi^2}{{l'}^2 \left( \varphi^2 + y_0^2 \right)}}{
\sinh^2 \left( \frac{\sqrt{y_0^2 + z_0^2} - \sqrt{y_0^2 + \varphi^2}}{l'} \right)} 
 - \frac{y_0^2}{l' \left( \varphi^2 + y_0^2 \right)^{\frac{3}{2}}}
\frac{ \cosh \left( \frac{\sqrt{y_0^2 + z_0^2} - \sqrt{y_0^2 + \varphi^2}}{l'} \right)}{
\sinh \left( \frac{\sqrt{y_0^2 + z_0^2} - \sqrt{y_0^2 + \varphi^2}}{l'} \right)} \right) \nn
&& - \frac{d(d-1)}{2} \frac{\varphi^2}{{l'}^2 \left( \varphi^2 + y_0^2 \right)}
\frac{ \cosh^2 \left( \frac{\sqrt{y_0^2 + z_0^2} - \sqrt{y_0^2 + \varphi^2}}{l'} \right)}{
\sinh^2 \left( \frac{\sqrt{y_0^2 + z_0^2} - \sqrt{y_0^2 + \varphi^2}}{l'} \right)} \nn
&& + \frac{d-1}{2l^2 \sinh^2 \left( \frac{\sqrt{y_0^2 + z_0^2} - \sqrt{y_0^2 + \varphi^2}}{l'} \right)}
\, .
\eea
Here we have identified $z$ as $\varphi$. 
We should note that $\omega(\varphi)$ is not always positive and hence the scalar field becomes ghost. 

We now show that we can construct a model without ghost. 
For simplicity, we only examine 
the case $D=d+1=5$. 
Let us explore 
the following model:
\be
\label{dSf1}
\e^{u(y)} = \frac{y^2}{l^2} \left( 1 + \frac{y^4}{y_0^4} \right)^{-1}\, .
\ee
When $y\to 0$, $\e^{u(y)}$ behaves as $\e^{u(y)} \sim y^2/l^2$. 
Therefore we can regard $y$ as a radial coordinate and $y=0$ corresponds to 
the center of the manifold whose topology of the spatial part is a four dimensional sphere. 
Since $d\left( \e^{u(y)} \right) |_{y=y_0} = 0$, we can cut the manifold at $y=y_0$ and glue 
two copies of the manifold cut at $y=y_0$. 
For the model (\ref{dSf1}), we find
\bea
\label{dSf2}
\omega(\varphi) &=& 3 \left( 1 + \frac{\varphi^4}{y_0^4} \right)^{-2} \frac{\varphi^2}{y_0^4} 
\left( 5 + \frac{\varphi^4}{y_0^4} \right) \left( 1 - \frac{\varphi^4}{y_0^4} \right)\, ,\nn
\mathcal{V}(\varphi) &=& \left( 1 + \frac{\varphi^4}{y_0^4} \right)^{-2} \left( \frac{75 \varphi^2}{2 y_0^4} 
+ \frac{6 \varphi^6}{y_0^8} + \frac{9 \varphi^{10}}{2y_0^{12}} \right)\, .
\eea
When $y^2=\varphi^2 \leq y_0^2$, both 
$\omega(\varphi)$ and $\mathcal{V}(\varphi)$ are positive 
and therefore there does not appear ghost. 
In order to avoid the ghost, we may restrict the value of $\varphi$ to be $\varphi^2 \leq y_0^2$
The energy density is also given by
\be
\label{dSf3}
\rho(y) = \left( 1 + \frac{y^4}{y_0^4} \right)^{-2} 
\left( \frac{45 y^2}{y_0^4} 
+ \frac{3 y^{10}}{y_0^{12}} \right)\, ,
\ee
which vanishes at $y=0$ and localizes at $y=y_0$ 
Then we have constructed a model of the de Sitter domain wall without ghost. 

We may remark 
that, when we glue the two copies of the region $y^2\leq y_0^2$, 
the value of $d^3 u/dy^3$ becomes discontinuous at $y=y_0$ although the values of 
$u$, $u'$, and $u''$ are continuous. This means 
that the derivative of the curvatures 
with respect to $y$ and therefore the derivative of the energy density in (\ref{dSf3}) 
become discontinuous at $y=y_0$. 
We should note, however, that this discontinuities never conflicts with the Einstein 
equation and field equation since these equations do not contain the derivative higher 
than two.  

\section{Localization of gravity  \label{III}}

As executed in \cite{Randall:1999vf}, we now investigate 
the localization of the gravity. 
Here we restrict the consideration to 
the case of $d=4$ for simplicity. 
For this purpose, we examine 
the following perturbation
\be
\label{lg1}
g_{\mu\nu} = g^{(0)}_{\mu\nu} + h_{\mu\nu}\, .
\ee
Here $g^{(0)}_{\mu\nu}$ is the metric given in the previous sections 
by solving the Einstein equations, etc. 
We express the quantities given by $g^{(0)}_{\mu\nu}$ by using the suffix $(0)$ and 
we define the lowering and raising of the vector index by using $g^{(0)}_{\mu\nu}$ and $g^{(0)\, \mu\nu}$ 
like $h^\mu_{\ \nu} = g^{(0)\, \mu\rho} h_{\rho\nu}$. 

We have 
\bea
\label{lg2}
\sqrt{-g} &=& \sqrt{-g^{(0)}} \left( 1 + \frac{1}{2} h_\mu^{\ \mu} 
+ \frac{1}{8} \left( h_\mu^{\ \mu} \right)^2 
- \frac{1}{4} h_{\mu\nu} h^{\mu\nu}
+ \mathcal{O}\left(h^3\right) \right)\, , \nn
R &=& R^{(0)} - R^{(0)\, \mu\nu} h_{\mu\nu} + \nabla^{(0)\, \mu} \nabla^{(0)\, \nu} h_{\mu\nu}
 - {\nabla^{(0)}}^2 h_\mu^{\ \mu} \nn
&& + \frac{3}{4} \left( \nabla^{(0)\, \sigma} h_\rho^{\ \rho} \right) \nabla^{(0)\, \nu} h_{\sigma\nu} 
+ \frac{3}{4} h^{\mu\nu} \nabla^{(0)}_\mu \nabla^{(0)}_\nu h_\rho^{\ \rho} 
 - \frac{1}{2} \nabla^{(0)\, \mu} h_{\mu\lambda} \nabla^{(0)}_\nu h^{\nu\lambda} \nn
&& + \frac{1}{4} h^{\mu\nu} {\nabla^{(0)}}^2 h_{\mu\nu}
 - \frac{1}{4} \left( \nabla^{(0)\, \sigma} h_\rho^{\ \rho} \right) 
\left( \nabla^{(0)}_{\sigma} h_\mu^{\ \mu} \right) 
+ \frac{1}{2} R^{(0)}_{\mu\nu} h^{\mu\rho} h^\nu_{\ \rho} 
+ \frac{1}{2} R^{(0)\, \rho\mu\sigma\nu} h_{\rho\sigma} h_{\mu\nu} \nn
&& + \nabla^{(0)\, \mu} \left( - \frac{1}{4} h_{\kappa\nu} \nabla_\mu h^{\nu\kappa} 
 - \frac{1}{2} h_{\mu\kappa}\nabla^{(0)}_\rho h^{\rho\kappa} 
+ \frac{1}{4} h_{\mu\kappa} \nabla^{(0)\, \kappa} h_\rho^{\ \rho} \right) \nn
&& + \frac{1}{2} {\nabla^{(0)}}^2 \left( h_{\mu\nu} h^{\mu\nu} \right) 
+ \mathcal{O}\left(h^3\right)\, .
\eea
Since we are interested in the localization of the graviton, which is massless and a spin two particle, 
we now assume, 
\be
\label{lg4}
h_{0\mu}=h_{y\mu} = \nabla^{(0)\,i} h_{ij} = h_i^{\ j} = 0\, .
\ee
In the following, we only study 
$D=d+1=5$ case for simplicity. 
Thus 
the action is reduced as follows: 
\bea
\label{lg5}
&& \int d^5 x \sqrt{-g} \left[ \frac{R}{2\kappa^2} 
 - \frac{1}{2}\omega(\varphi) \partial_\mu \varphi \partial^\mu \varphi
 - \mathcal{V}(\varphi) \right] \nn
&& \to \int d^D x \sqrt{-g^{(0)}} \left[ 
\frac{1}{2\kappa^2} \left\{ R^{(0)} - \frac{1}{4} \nabla^{(0)\, \rho} h^{ij} \nabla^{(0)}_\rho h_{ij} 
+ \frac{1}{2} R^{(0)\, ij}h_{ik} h_j^{\ k} 
+ \frac{1}{2} R^{(0)\, ikjl} h_{ij} h_{kl} \right. \right. \nn
&& \qquad \left. \left. - \frac{1}{4} R^{(0)} h_{ij}h^{ij} \right\}
 - \frac{1}{4}h_{ij} h^{ij} \left( - \frac{1}{2}\omega(\varphi) \partial_\mu \varphi \partial^\mu \varphi
 - \mathcal{V}(\varphi) \right) + \mathcal{O} \left(h^3\right)
\right] \, .
\eea
Then by the variation of $h_{ij}$, we obtain
\bea
\label{lg6}
0 &=& \frac{1}{2\kappa^2} \left\{ \frac{1}{2} \nabla^\rho \nabla_\rho h_{ij} 
+ \frac{1}{2} \left( R^{(0)}_{ik} h^k_{\ j} + R^{(0)}_{jk} h^k_{\ i} \right) \right. \nn
&& \left. + R^{(0)}_{ikjl} h^{kl} - \frac{1}{2} R^{(0)} h_{ij} \right\} 
 - \frac{1}{2\kappa^2} h_{ij} \left( - \frac{1}{2} \omega\left( \varphi \right) 
\partial_\mu \varphi \partial^\mu \varphi - \mathcal{V}(\varphi) \right)\, .
\eea

First we explore 
the case that the domain wall is flat, i.e., 
${\hat g}_{\mu\nu}=\eta_{\mu\nu}$ in (\ref{I2}). 
Since 
\be
\label{lg7}
\nabla^\rho \nabla_\rho h_{ij} = \left( \partial_y^2 + \e^{-u}\Box \right) h_{ij} 
 - \left( u'' + \frac{3}{2} {u'}^2 \right) h_{ij}\, ,
\ee
we find 
\be
\label{lg8}
0 = \left( \partial_y^2 + \e^{-u} \Box \right) h_{ij} 
+ \left( - u'' - {u'}^2 \right) h_{ij}\, .
\ee
Here we have used (\ref{I5}) and (\ref{I6}) with $d=4$. 
We should note that there is always a zero-mode solution, where $\Box h_{ij} = 0$ 
in the flat (domain wall) space-time. 
The explicit form of the zero-mode is given by
\be
\label{lg9}
h_{ij} \propto \e^{u}\, ,
\ee
which could surely be normalizable if $u$ sufficiently rapidly goes to minus infinity 
$u\to + \infty$ when $|y|\to \infty$. 

Next we investigate 
the case that the domain wall is de Sitter or anti-de Sitter space-time, 
namely, 
${\hat R}_{\mu\nu} = \frac{3}{l^2} {\hat g}_{\mu\nu}$. 
Then instead of (\ref{lg8}), we obtain
\be
\label{lg10}
0 = \left( \partial_y^2 + \e^{-u} \Box \right) h_{ij} 
+ \left( - u'' - {u'}^2 - \frac{2\e^{-u}}{l^2}\right) h_{ij}\, .
\ee
Here we have used (\ref{I5}) and (\ref{I6}) with $d=4$. 
In the (anti)-de Sitter space-time, the zero-mode solution, which corresponds to the 
massless graviton, is given by 
\be
\label{lg11}
\Box h_{ij} = \frac{2}{l^2} h_{ij}\, .
\ee
Therefore the zero-mode is again given by (\ref{lg9}). 
Accordingly 
the zero-mode is always proportional to the warp factor $\e^{u}$. 
As a result, 
even in the (anti-)de Sitter domain wall, there occurs the localization of graviton.  

\section{Reconstruction of general FRW domain wall universe \label{IIIb}} 

In this section, as an extension of the previous sections, 
we examine more general domain wall, which can be regarded as a general FRW universe. 
The metric we are considering is given by 
\begin{align}
\label{metric}
ds^2 &= dw^2 + f \left( w, t \right) \left\{   \frac{dr^2}{1-kr^2} + r^2 d \theta^2 
+r^2 \sin^2 \theta d \phi^2  \right\} - \frac{e\left( w,t \right)^2}{f\left( w,t \right)} dt^2 
\nonumber \\
& \equiv \e^{\ln f \left( w,t \right)} \gamma_{mn} \left( x \right) dx^m dx^n 
+ h_{\alpha \beta} \left( y \right) dy^{\alpha} dy^{\beta} \, .
\end{align}
Here $m,n = 1,2,3$, $\alpha,\beta = 0,5$, and $y^0 = t$, $y^5= w$. 
Especially if we choose
\be
\label{FRWchoice}
f \left( w,t \right) = L^2 \e^{u \left( w,t \right)} a \left( t \right)^2 \, ,\quad 
e \left( w, t \right) = L^2 \e^{u \left( w,t \right)} a \left( t \right) \, ,
\ee
the general FRW universe, whose metric is described by
\be
\label{FRWmetric0}
ds_\mathrm{FRW}^2 = - dt^2 + a \left( t \right)^2 \left\{   \frac{dr^2}{1-kr^2} + r^2 d \theta^2 
+r^2 \sin^2 \theta d \phi^2  \right\}\, ,
\ee
is embedded by the arbitrary warp factor $L^2 \e^{u \left( w,t \right)}$. 

For the metric (\ref{metric}), the connections can be given by
\begin{align}
\label{connection1}
& \hat{\Gamma}^m_{\ ab} = \Gamma^m_{\ ab} \left( \gamma \right)\, ,\quad 
\hat{\Gamma}^m_{\ n \alpha} = \hat{\Gamma}^m_{\ \alpha n} 
= \frac{1}{2} \delta^m_n \frac{\partial_\alpha f}{f}\, ,\quad 
\hat{\Gamma}^{\alpha}_{\ \beta \gamma} = \Gamma^{\alpha}_{\ \beta \gamma} \left( h \right)\, ,\quad 
\hat{\Gamma}^{\alpha}_{\ mn} = - \frac{1}{2} \gamma_{mn} h^{\alpha \beta} \partial_{\beta} f\, ,
\end{align}
that is,
\begin{align}
\label{connection2}
& \Gamma^{1}_{\ 11} = \frac{kr}{1-kr^2}\, ,\quad 
\Gamma^{2}_{\ 12} = \frac{1}{r}\, ,\quad 
\Gamma^{3}_{\ 13} = \frac{1}{r}\, ,\quad 
\Gamma^{1}_{\ 22} = -r \left( 1 - k r^2  \right)\, ,\nonumber \\
& \Gamma^{3}_{\ 23} = \cot \theta\, ,\quad 
\Gamma^{1}_{\ 33} = -r \left( 1 - k r^2  \right) \sin^2 \theta\, ,\quad 
\Gamma^{2}_{\ 33} = - \cos \theta \sin \theta \, ,\nonumber \\
& \Gamma^{0}_{\ 00} = - \frac{e \dot{f} - 2 \dot{e} f}{2ef} \, ,\quad 
\Gamma^{5}_{\ 00} = - \frac{e^2 f^\prime - 2 e e^\prime f}{2 f^2}\, ,\quad 
\Gamma^{0}_{\ 05} = - \frac{e f^\prime -2 e^\prime f}{2 e f}\, .
\end{align}
Then the non-vanishing Riemann tensors are expressed by
\begin{align}
\label{curvature1}
\hat{R}^{a}_{\ bcd } &=  {}^{\left( 3 \right) }R^{a}_{\ bcd} \left( \gamma \right)
- \frac{1}{4} \left(  \delta^a_c \gamma_{db} - \delta^a_d \gamma_{c b} \right) 
\left\{  \left( f^\prime  \right)^2 - \left( \frac{\dot{f}}{e^2}  \right)^2   \right\}\, ,\nonumber \\
\hat{R}^{a}_{\ \gamma b \delta} &= - \frac{1}{2} \delta^{a}_{b} \nabla^{\left( h \right)}_{\delta} 
\nabla^{\left( h \right)}_{\gamma} \ln f - \frac{1}{4} \delta^{a}_{b} 
\nabla^{\left( h \right)}_\gamma \ln f \nabla^{\left( h \right)}_\delta \ln f \, , \nonumber \\
\hat{R}^{\alpha}_{\ b c \delta} &= \frac{1}{2} \gamma_{c b} \nabla^{\left( h \right)}_\delta 
\left(  h^{\alpha \kappa} \nabla^{\left( h \right)}_\kappa f   \right) 
 - \frac{1}{4} g_{c b} f h^{\alpha \kappa} \frac{\partial_\kappa f \partial_\delta f}{f}\, ,\nonumber \\
\hat{R}^{\alpha}_{\ \beta \gamma \delta} & ={}^{\left( 2 \right)} 
R^{\alpha}_{\ \beta \gamma \delta} \left( h \right)\, ,
\end{align}
where
\begin{align}
\label{curvature2}
& {}^{\left( 3 \right)} R^{2}_{\ 121} = {}^{\left( 3 \right)} R^{3}_{\ 131}  =\frac{k}{1- k r^2} \, ,\quad 
{}^{\left( 3 \right)} R^{1}_{\ 221} = - {}^{\left( 3 \right)} R^{3}_{\ 232} = - k r^2 \, ,\quad 
{}^{\left( 3 \right)} R^{1}_{\ 331} = {}^{\left( 3 \right)} R^{2}_{\ 332} = - k r^2 \sin^2 \theta \, ,
\nonumber \\
& {}^{\left( 2 \right)} R^{5}_{\ 050} = \frac{e^2}{f} {}^{\left( 2 \right)} R^{0}_{\ 550} 
= - \frac{  2 e^2 f f^{\prime \prime} - 3 e^2 \left( f^{\prime}  \right)^2 
+ 4 e e^\prime f f^\prime - 4e e^{\prime \prime } f^2  }{4 f^3}\, .
\end{align}
Then the Einstein tensors have the following forms:
\begin{align}
G^{0}_{\ 0} &= \frac{3f^{\prime \prime}}{2f} - \frac{3k}{f} - \frac{3 \dot{f}^2 }{4e^2 f} \, , \nonumber \\
G^{1}_{\ 1} &= G^{2}_{\  2} = G^{3}_{\ 3} 
= \frac{f^{\prime \prime}}{2f} + \frac{e^{\prime \prime}}{e} - \frac{k}{f} - \frac{\ddot{f}}{e^2} 
 - \frac{\dot{f}^2}{4e^2 f} + \frac{\dot{e} \dot{f}}{e^3} \, ,\nonumber \\
G^{5}_{\ 5} &= \frac{3 f^{\prime} e^{\prime}}{2fe} - \frac{3k}{f} - \frac{3 \ddot{f}}{2e^2} 
 - \frac{3 \dot{f}^2}{4 e^2 f} + \frac{3 \dot{e} \dot{f}}{2e^3} \, ,\nonumber \\
G^{5}_{\ 0} &= - \frac{3 e^{\prime} \dot{f}}{2 e^3} + \frac{3 \dot{f}^{\prime}}{2 e^2} \, .
\end{align}

We now investigate the action of the scalar fields $\phi$ and $\chi$:
\be
\label{pc1}
S_{\phi\chi} = \int d^4 x \sqrt{-g} \left\{ - \frac{1}{2} A (\phi,\chi) \partial_\mu \phi \partial^\mu \phi 
 - B (\phi,\chi) \partial_\mu \phi \partial^\mu \chi 
 - \frac{1}{2} C (\phi,\chi) \partial_\mu \chi \partial^\mu \chi - V (\phi,\chi)\right\}\, .
\ee
We here construct a model to realize the arbitrary metric which can be written in the form of (\ref{metric}). 

For the model (\ref{pc1}), the energy-momentum tensor could be given by
\begin{align}
\label{pc2}
T^{\phi\chi}_{\mu\nu} =& g_{\mu\nu} \left\{ 
 - \frac{1}{2} A (\phi,\chi) \partial_\rho \phi \partial^\rho \phi 
 - B (\phi,\chi) \partial_\rho \phi \partial^\rho \chi 
 - \frac{1}{2} C (\phi,\chi) \partial_\rho \chi \partial^\rho \chi - V (\phi,\chi)\right\} \nn
& +  A (\phi,\chi) \partial_\mu \phi \partial_\nu \phi 
+ B (\phi,\chi) \left( \partial_\mu \phi \partial_\nu \chi + \partial_\nu \phi \partial_\mu \chi \right) 
+ C (\phi,\chi) \partial_\mu \chi \partial_\nu \chi \, .
\end{align}
On the other hand, the field equations read
\begin{align}
\label{pc3}
0 =& \frac{1}{2} A_\phi \partial_\mu \phi \partial^\mu \phi + A \nabla^\mu \partial_\mu \phi 
+ A_\chi \partial_\mu \phi \partial^\mu \chi 
+ \left( B_\chi - \frac{1}{2} C_\phi \right)\partial_\mu \chi \partial^\mu \chi  
+ B \nabla^\mu \partial_\mu \chi - V_\phi \, ,\\
\label{pc4}
0 =& \left( - \frac{1}{2} A_\chi + B_\phi \right) \partial_\mu \phi \partial^\mu \phi 
+ B \nabla^\mu \partial_\mu \phi 
+ \frac{1}{2} C_\chi \partial_\mu \chi \partial^\mu \chi 
+ C \nabla^\mu \partial_\mu \chi + C_\phi \partial_\mu \phi \partial^\mu \chi 
 - V_\chi\, .
\end{align}
Here $A_\phi=\partial A(\phi,\chi)/\partial \phi$, etc. 
We now choose $\phi=t$ and $\chi=w$. Then we find
\be
\label{pc4b}
T_0^{\ 0} = - \frac{f}{2e^2} A - \frac{1}{2} C - V\, ,\quad 
T_i^{\ j} = \delta_i^{\ j} \left( \frac{f}{2e^2} A - \frac{1}{2} C - V \right)\, ,\quad 
T_5^{\ 5} = \frac{f}{2e^2} A + \frac{1}{2} C - V\, ,\quad 
T_0^{\ 5} = B \, ,
\ee
and 
\begin{align}
\label{pc5}
0 =& - \frac{f}{2e^2} A_\phi + \frac{f}{e^2} \left( \frac{\dot e}{e} - \frac{2\dot f}{f} \right)A 
+ B_\chi + B \left( \frac{e'}{e} + \frac{f'}{f} \right) - \frac{1}{2} C_\phi - V_\phi \, ,\\
\label{pc9}
0 =& \frac{f}{2e^2} A_\chi - \frac{f}{e^2} B_\phi 
+ \frac{f}{e^2} \left( \frac{\dot e}{e} - \frac{2\dot f}{f} \right) B
+ \frac{1}{2} C_\chi + C \left( \frac{e'}{e} + \frac{f'}{f} \right) - V_\chi \, .
\end{align}
Eqs.~(\ref{pc4b}) can be solved with respect to $A$, $B$, $C$, and $V$ as follows, 
\begin{align}
\label{pc7}
A =& \frac{e^2}{\kappa^2 f} \left( G_1^{\ 1} - G_0^{\ 0} \right) 
=  \frac{e^2}{\kappa^2 f} \left( G_2^{\ 2} - G_0^{\ 0} \right) 
=  \frac{e^2}{\kappa^2 f} \left( G_3^{\ 3} - G_0^{\ 0} \right) 
\nn
=& \frac{1}{\kappa^2} \left( - \frac{e^2 f''}{f^2} + \frac{e e''}{f} + \frac{2ke^2}{f^2} 
 - \frac{\ddot f}{f} + \frac{{\dot f}^2}{2 f^2} + \frac{\dot e \dot f}{ef} \right) \, , \nn
B =& \frac{1}{\kappa^2}G_0^{\ 5} = \frac{1}{\kappa^2} 
\left( - \frac{3 e' \dot f}{2 e^3} + \frac{3{\dot f}'}{2e^2} \right) \, , \nn
C =& \frac{1}{\kappa^2} \left( G_5^{\ 5} - G_1^{\ 1} \right)
= \frac{1}{\kappa^2} \left( G_5^{\ 5} - G_2^{\ 1} \right) 
= \frac{1}{\kappa^2} \left( G_5^{\ 5} - G_3^{\ 1} \right) \nn
=& \frac{1}{\kappa^2} \left( - \frac{f''}{2f} - \frac{e''}{e} - \frac{2k}{f} 
 - \frac{\ddot f}{2e^2} - \frac{{\dot f}^2}{2e^2 f} 
+ \frac{\dot e \dot f}{2e^3} + \frac{3 f' e'}{2fe} \right)\, , \nn
V =& \frac{1}{\kappa^2} \left( G_0^{\ 0} + G_5^{\ 5} \right) \nn
& \frac{1}{\kappa^2} \left( - \frac{3f''}{4f} + \frac{3k}{f} 
+ \frac{3 {\dot f}^2}{4e^2 f} 
- \frac{3f'e'}{4fe} + \frac{3\ddot f}{4e^2} - \frac{3\dot e \dot f}{4e^3}
\right)\, .
\end{align}
Then we find that the field equations (\ref{pc5}) and (\ref{pc9}) are nothing but the 
Bianchi identities:
\bea
\label{pc8}
&& - \frac{f}{2e^2} A_\phi + \frac{f}{e^2} \left( \frac{\dot e}{e} - \frac{2\dot f}{f} \right)A 
+ B_\chi + B \left( \frac{e'}{e} + \frac{f'}{f} \right) - \frac{1}{2} C_\phi - V_\phi 
= - \frac{e^2}{2f} \nabla^\mu G_\mu^{\ 0}\, ,\\
\label{pc10}
&& \frac{f}{2e^2} A_\chi - \frac{f}{e^2} B_\phi 
+ \frac{f}{e^2} \left( \frac{\dot e}{e} - \frac{2\dot f}{f} \right) B
+ \frac{1}{2} C_\chi + C \left( \frac{e'}{e} + \frac{f'}{f} \right) - V_\chi 
= \nabla^\mu G_\mu^{\ 5}\, .
\eea
Therefore the field equations (\ref{pc5}) and (\ref{pc9}) are surely satisfied by choosing 
$A$, $B$, $C$, and $V$ by (\ref{pc7}). 

Thus we can construct a model, where the general FRW universe can be embedded in an arbitrary 
warp factor, by the choice of $A$, $B$, $C$, and $V$ in (\ref{pc7}). 

\section{Examples of reconstructed model}

In this section, we show some examples of reconstruction by using two scalar fields, 
where no ghost field appears. 
For simplicity, we consider examples in case $k=0$.

As a first example, we assume $a(t) \propto t^{h_0}$ with a constant $h_0$. 
Then, the equations in (\ref{pc7}) give,
\bea
\label{U1}
\kappa^2 A &=& - \frac{\ddot U}{U} + \frac{3{\dot U}^2}{2 U^2} + \frac{h_0 \dot U}{t U} 
+ \frac{2h_0}{t^2}\, , \nn
L^2 \kappa^2 B &=& - \frac{3U' \dot U}{2 U^3} + \frac{3 {\dot U}'}{2U^2} \, , \nn
\kappa^2 C &=& - \frac{3 U''}{2U} + \frac{3 {U'}^2}{2 U^2} 
+ \frac{1}{L^2}\left( - \frac{\ddot U}{2 U^2} - \frac{5h_0 \dot U}{2 t U^2} 
 - \frac{3h_0^2 - h_0}{t^2U} \right)\, .
\eea
Here $U \left( w,t \right) \equiv\e^{u \left( w,t \right)}$. 
Furthermore by assuming $U \left( w,t \right) = W\left( w \right) T \left(t\right)$, we rewrite 
(\ref{U1}) as follows,
\bea
\label{U2}
\kappa^2 A &=& - \frac{\ddot T}{T} + \frac{3{\dot T}^2}{2 T^2} 
+ \frac{h_0 \dot T}{t T} + \frac{2h_0}{t^2}\, , \nn
L^2 \kappa^2 B &=& 0 \, , \nn
\kappa^2 C &=& - \frac{3 W''}{2W} + \frac{3 {W'}^2}{2 W^2} 
+ \frac{1}{L^2 WT^2}\left( - \frac{\ddot T}{2} - \frac{5h_0 \dot T}{2 t}
 - \frac{\left( 3h_0^2 - h_0\right) T}{t^2}\right)\, .
\eea
If we assume $T \propto t^\beta$, we find
\begin{align}
\label{U2B}
 - \frac{\ddot T}{2} - \frac{5h_0 \dot T}{2 t}
 - \frac{\left( 3h_0^2 - h_0\right) T}{t^2}
&\propto - \frac{1}{2} \left( \beta^2 - \left(1 - 5h_0 \right) \beta + 6 h_0^2 - 2 h_0 \right)  \nn
&= -\frac{1}{2} \left\{  \beta + \left(3 h_0 -1 \right) \right\} 
\left\{ \beta + 2 h_0 \right\} \, .
\end{align}
Then $T$ is given by
\be
\label{U3}
T(t) = T_1 t^{1 - 3 h_0} + T_2 t^{- 2h_0}\, .
\ee
Therefore we obtain
\be
\label{U5}
A = \frac{1}{\kappa^2} \left\{ \frac{3}{2}\left( \frac{\dot T(t)}{T(t)}
+ \frac{2h_0}{t} \right)^2  \right\}>0\, ,\quad 
\kappa^2 C = - \frac{3 W''}{2W} + \frac{3 {W'}^2}{2 W^2} \, .
\ee
We may choose
\be
\label{U6}
W(w) = \e^{- \frac{w^2}{w_0^2}}\, ,
\ee
with a constant $w_0$, 
then we find 
\be
\label{U7}
C = \frac{3}{\kappa^2 w_0^2}>0 \, .
\ee
Because both of $A$ and $C$ are positive and $B$ vanishes, any ghost does not appear 
in this model. 
Since $a(t) \propto t^{h_0}$,  the domain universe corresponds 
to the universe filled with the perfect fluid whose equation of state 
parameter $w$ is given by 
\be
\label{AAA3}
w = -1 + \frac{2}{3h_0}\, .
\ee
Since we now have $f \left(  w,t \right) = L^2 T \left( t \right) \e^{- \omega^2 / \omega^2_0} a^2_0 t^{2h_0}$, 
$e \left(  w,t \right) = L^2 T \left( t \right) \e^{- \omega^2 / \omega^2_0} a_0 t^{h_0}$ in 
(\ref{FRWchoice}). 
Then by using the last equation in (\ref{pc7}), we find the explicit form of the potential $V$:
\begin{align}
\label{AAAA1}
\kappa^2 V \left( w,t \right) = - \frac{3}{4} \left[ -2 + 8w^2 
+ \left( \frac{\dot{T}}{T} + \frac{2 h_0}{t} \right)^2 a_0 t^{h_0} 
+ \frac{\e^{-w^2 / w^2_0}}{L^2 T }  \left(  \frac{\dot{T}^2}{T^2} 
 - \frac{\ddot{T}}{T} - \frac{h_0 \dot{T}}{t T} \right) \right]\, .
\end{align}
By replacing $t$ and $w$ by $\phi$ and $\chi$, we obtain the explicit form 
of the potential $V \left( \phi , \chi  \right)$ 
in terms of the scalar fields $\phi$ and $\chi$.

As another example, we may consider the de Sitter universe, $a \left( t \right) = a_0 \e^{H_0 t}$ 
with constants $a_0$ and $H_0$. 
As we saw in the previous sections, the de Sitter universe can be realized even in one scalar model 
but we now consider this example as a simple demonstration of the reconstruction by using two 
scalar fields. 
Because 
\be
\label{AAAA2}
f \left( w, t \right) = L^2 \e^{u(w,t)} a^2_0 \e^{2 H_0 t} \, \quad 
e \left( w, t \right) = L^2 \e^{u (w,t)} a_0 \e^{H_0 t} \, ,
\ee
we find
\begin{align}
\label{AAAA3}
\kappa^2 A \left( w,t  \right) &= \frac{1}{2} \dot{u}^2 + H_0 \dot{u} + H^2_0 - \ddot{u}\, , \\
\label{AAAA4}
\kappa^2 B \left( w,t  \right) &= \frac{3}{2} L^{-2} \e^{-u} \dot{u}^{\prime}\, , \\
\label{AAAA5}
\kappa^2 C \left( w,t  \right) &= - \frac{3}{2} u^{\prime \prime} 
 - \frac{\e^{-u}}{2L^2 } \left\{ \dot{u}^2 + 5H_0 \dot{u} + 6H^2_0 + \ddot{u}  \right\}\, .
\end{align}
The second term of the r.h.s. in (\ref{AAAA5}), which is proportional to $L^{-2}$ vanishes, if we choose 
\be
\label{AAAA6}
u \left( w,t  \right) = -3 H_0 t + \log \left( - \alpha \left( w \right) 
+ \e^{H_0 t}  \right) + \beta \left( w \right) \, ,
\ee
where $\alpha \left( w \right) $ and $\beta \left( w \right)$ are arbitrary functions 
which does not depend on $t$ but only depend on $w$.
If we substitute (\ref{AAAA6}) into the expressions of $A$, $B$, and $C$ 
in (\ref{AAAA3}), (\ref{AAAA4}), and (\ref{AAAA5}), respectively, we find 
\begin{align}
\label{AAAA7}
\kappa^2 A \left( w,t  \right) &= \frac{  H^2_0}{ \left( \e^{H_0 t} 
 - \alpha \right)^2} \left\{ \e^{2 H_0 t}   - \frac{\alpha^2}{2}  \right\} \, ,\\
\label{AAAA8}
\kappa^2 B \left( w,t  \right) &= \frac{3}{2} L^{-2} \e^{-u} 
\frac{H_0 \alpha^{\prime} \e^{H_0 t}}{ \left( \e^{H_0 t} - \alpha  \right)^2} \, ,\\
\label{AAAA9}
\kappa^2 C \left( w,t  \right) &= - \frac{3}{2} \frac{ \alpha^{\prime \prime } 
\left( w \right) \e^{H_0 t} - \alpha \left( w \right) \alpha^{\prime \prime} 
\left( w \right) + \left(  \alpha^{\prime} \left( w \right) \right)^2 }
{ \left( \e^{H_0 t} - \alpha  \right)^2 } - \frac{3}{2} \beta^{\prime \prime} 
\left( w \right)\, .
\end{align}
Then by choosing $\alpha = 0$ and $\beta^{\prime \prime} <0 $, we find 
\begin{align}
\label{AAAA10}
\kappa^2 A \left( w,t  \right) &= H^2_0 \, ,\\
\label{AAAA11}
\kappa^2 B \left( w,t  \right) &= 0 \, ,\\
\label{AAAA12}
\kappa^2 C \left( w,t  \right) &= -\frac{3}{2} 
\beta^{\prime \prime } \left( w \right)\, .
\end{align}
Because $A$ and $C$ are positive definite, no ghost field appears in this model. 

Thus, we reconstruct two models, corresponding to the power-law expanding universe 
and to the de Sitter universe. The latter has been constructed by using one scalar model as 
in the previous section. As we saw, however, the argument about the existence of ghost becomes 
rather simpler in the two scalar model than in the one scalar model. 
By smoothly connecting the two models corresponding to the power-law expanding universe 
and to the de Sitter universe, we may construct a model which unifies, inflation, matter dominant 
universe, and the late-time accelerating expansion of the universe.

\section{Summary and discussions \label{IV}}

By using the formulation of the reconstruction, we have found the models which have an exact 
solution describing the domain wall. The shape of the domain wall can be flat, 
de Sitter space-time, or anti-de Sitter space-time.  
In the domain wall solutions, there often appears ghost with negative kinetic energy. 
We have constructed, however, 
an example of the de Sitter domain wall solution without ghost, which can be a toy model 
of inflation. We have also investigated the localization of gravity and it has been 
demonstrated that the four dimensional Newton law could be reproduced. 

We have also shown that a space-time, where the domain wall is the general FRW 
universe and the warp factor can be arbitrary, can be constructed by using the 
two scalar fields. 
It has been shown that the scalar field equations are equivalent to the Bianchi identities: 
$\nabla^\mu \left( R_{\mu\nu} - \frac{1}{2}R g_{\mu\nu} \right)=0$.

We have not, however, studied 
if the domain wall solution is stable or unstable. 
About the previous work on the stability of the domain wall, see \cite{Kobayashi:2001jd}. 

For the check of the (in)stability, we need to consider the time-dependent perturbation 
from the solution. The existence of the massless graviton, which is obtained from 
the fluctuation of the metric, may inform 
that the model could be stable under the perturbation 
of the metric. In order to verify the stability, however, we of course need to include the 
perturbation of the scalar field $\varphi$, or 
$\phi$ and $\chi$, which could be a future work. 

\section*{Acknowledgments.}

We are grateful to S.~D.~Odintsov for the discussion when he stays in 
Nagoya University. 
S.N. is supported by Global COE Program of Nagoya University (G07)
provided by the Ministry of Education, Culture, Sports, Science \&
Technology and by the JSPS Grant-in-Aid for Scientific Research (S) \# 22224003
and (C) \# 23540296.

\end{document}